# Performance Analysis of Pulse Shaping Technique for OFDM PAPR Reduction

S. M. Kamruzzaman and Md. Anisur Rahman

Department of Information and Communication Engineering
University of Rajshahi, Rajshahi-6205, Bangladesh.
E-mail: smzaman@gmail.com, anisur_81@yahoo.com

**Abstract.** Orthogonal Frequency Division Multiplexing (OFDM) is an attractive modulation and multiple access techniques for channels with a nonflat frequency response, as it saves the need for complex equalizers. It can offer high quality performance in terms of bandwidth efficiency, robustness against multipath fading and cost-effective implementation. However, its main disadvantage is the high peak-to-average power ratio (PAPR) of the output signal. As a result, a linear behavior of the system over a large dynamic range is needed and therefore the efficiency of the output amplifier is reduced. In this paper, we investigate the effect of some of these sets of time waveforms on the OFDM system performance in terms of Bit Error Rate (BER). We evaluate the system performance in AWGN channels. The obtained results indicate that the reduction in PAPR of the investigated methods is associated with considerable improvement in BER performance of the system, in multipath channels, as compared to conventional OFDM. These promising results indicate that pulse shaping with reduced PAPR is an attractive solution for an OFDM system.

**Keywords:** OFDM, PAPR, sub-carriers, pulse shaping, multipath channels, AWGN channels, BER.

## 1. Introduction

Wireless digital communications is rapidly expanding resulting in a demand for wireless systems that are reliable and have a high spectral efficiency [7] [8]. Orthogonal frequency division multiplexing (OFDM) is becoming the chosen modulation technique for wireless communications. OFDM can provide large data rates with sufficient robustness to radio channel impairments. Recently, OFDM systems are being applied for fixed and mobile transmission [9]. Some examples of existing systems where OFDM is used are digital audio and video broadcasting, and Asymmetric Digital Subscriber Line (ADSL) modems. These support a physical layer transmission rate up to 54 Mbps and use OFDM for the physical layer implementation. Additionally, OFDM is being considered for future broadband applications such as Wireless Asynchronous Transfer Mode (WATM) and fourth generation (4G) transmission techniques.

There are two main drawbacks with OFDM [6], the large dynamic range of the signal also referred as peak-to average power ratio (PAPR) and its sensitivity to frequency errors. These in turn are the main research topics of OFDM in many research centers around the world. Minimizing the PAPR allows a higher average power to be transmitted for a fixed peak power, improving the overall signal to noise ratio at the receiver [10] [11]. It is therefore important to minimize the PAPR.

## 2. Choice of PAPR Reduction Technique

There have been many methods in which the power ratio can be reduced in OFDM multi-carrier signals. Some researchers have developed methods of achieving low PAPR through the application of "clipping and filtering methods, block coding and selected mapping" [1] [2]. Other techniques of lowering the PAPR ratio, which have been published, involve complicated applications of supplementary FFT function blocks requiring more functional components [12]. In addition, non-hardware based approaches that also incur other disadvantages such as adding information in overhead parameters or handshaking algorithms to control PAPR levels [3]. Additional information bearing causes OFDM signal to be less bandwidth efficient [4] [5].

## 3. Crosscorrelation

From MATLAB simulation, the following normalized crosscorrelation plot was obtained shown in Fig. 1. A generalization can be made from the plot which shows the steady improvement in the crosscorrelation between pulse samples for a gradual increase of 'n'. The relationship can be established that the characteristics of increasing main lobe width and decreasing sidelobe peaks are favorable for improvements in crosscorrelation. The distance between the zero crossings of the attenuations apparent in the red plot represents the 1/T separation. Comparing the cut-off frequency of the crosscorrelation with respect to parameter 'n', there is an increase by a factor of 4 in the cut-off frequency of the main lobe comparing n = 4 to n = 0. Similarly for n = 16, there is roughly a factor 2 increase in the cut-off frequency over the previous plot for n = 4.

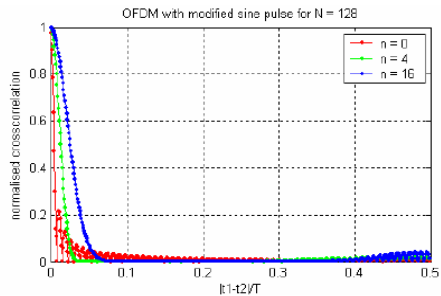 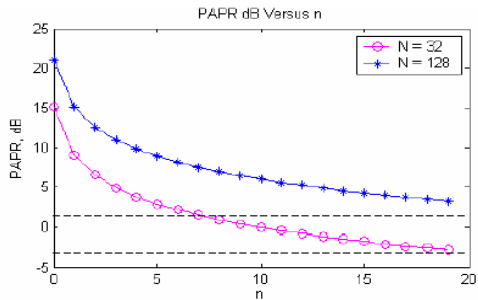

**Fig. 1.** Crosscorrelation Plot (Sine Pulse).   **Fig. 2.** Maximum PAPR (Normalized Sine Pulse).

The enhancement level of the crosscorrelation start to reduce for large 'n' and this has been illustrated by the crosscorrelation cut-off frequencies. In MATLAB, the improvements of the crosscorrelation are negligible for high 'n' for example beyond n = 40. This behavior is more clearly demonstrated when the maximum PAPR is plot against shape parameter 'n'.

### 3.1 Maximum PAPR Plot

For the normalized sine pulse case, the plot of the maximum PAPR against the shape parameter 'n' is illustrated in Figs. 3-4. By plotting the maximum PAPR for a normalized sine pulse, an attractive result was gained. It can be observed that the normalization of sine pulses have brought PAPR levels lower than the unmodified sine pulse.

The maximum PAPR has changed considerably achieving a greater rate of PAPR reduction per 'n'. This implies that lowering the energy of the pulses has direct impact on the lowering of PAPR. One important factor to note in Fig. 1 and 2 is that when number of subcarriers N is increased in the system, the PAPR increases dramatically. This behavior is coincides with the theory of high PAPR in OFDM signals. The cause of high PAPR in OFDM signals is due to the separation of transmissions into many subcarriers. Due to the composition of many subcarriers, the sums of all subcarrier signals lead to constructive superposition creating regions of large power peaks, thus contributing to high PAPR.

## 4. Experimental Results

Fig. 3 shows the overall performance of OFDM communication system. The simulated BER always highly exceeds the theoretical BER. In this case, larger the BER, more degrades the performance.

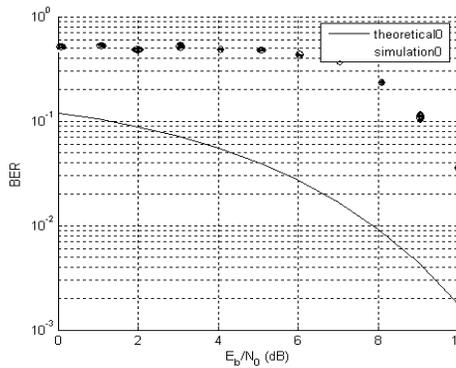 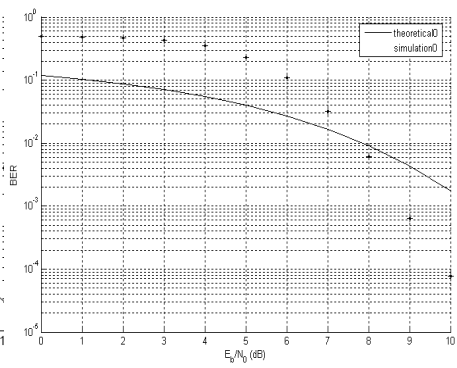

**Fig.3.** Simulated and theoretical Bit Error Rate (BER).  **Fig. 4**. Simulated and theoretical Bit Error Rate (BER).

### 4.1 BER using Pulse Shaping

Numerical results indicate a direct relationship between the width of the main lobe pulse response and the cut-off frequency of the crosscorrelation plots. Intuitively, wider main lobes induce larger crosscorrelation between subcarriers. The worst case PAPR values are obtained through the use of rectangular subcarrier pulses with high sidelobe peaks in their response. It is suggested that pulses with relatively low side lobe peaks should be used to deviate the worst-case scenario. This is because the worst case PAPR is obtained by using rectangular pulses. From broadband pulse tests, flat main lobe responses and rapid pass band to stop band transitions increase the area of the crosscorrelation plots. Thus, better correlation is achieved between subcarriers if the carriers possess all four main pulse characteristics, effectively reducing the PAPR. The pulse shaping technique proves to be a simple approach to PAPR reduction based on modifications to the existing OFDM signal structure.

### 4.1 Simulation Results

Fig. 3 shows the overall performance of OFDM communication system. The pulse shaping technique has been used to test narrowband and broadband pulses concluding that broadband pulse were far superior in reducing PAPR. From the graph, the simulated BER is greater than the theoretical BER at the low value of Eb/No, But when this value increases then the BER value significantly decreases than the theoretical BER value.

### 4.1.1 BER for different M of M-ary QAM

Fig. 5 shows the effect of BER without using pulse shaping for M=4, 8, 16, 32. The lower the value of M, better the value of BER. From Fig. 6, we observe that the lower the value of M, better the value of BER. But in this case, the value of BER decreases significantly compared to the former.

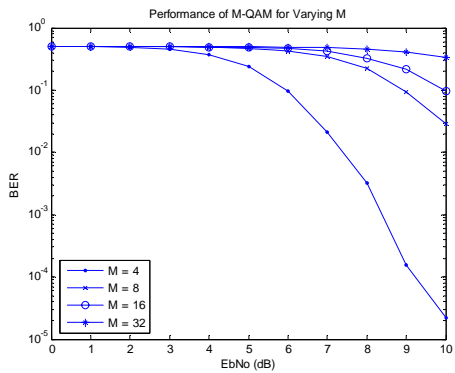
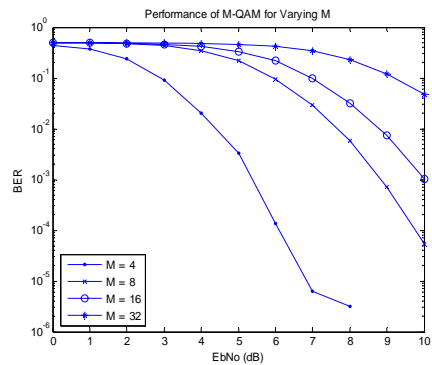

**Fig 5.** Simulated Bit Error Rate for M=4, 8, 16, 32 (without using pulse shaping).

**Fig. 6.** Simulated Bit Error Rate for M=4, 8, 16, 32 (using pulse shaping).

## 5. Conclusion

In this paper, a problem experienced in the OFDM modulation technique has been studied. The pulse shaping technique used in this work performs well in assisting the study of efficient pulse characteristics that improve PAPR reduction. Four subcarrier pulse shapes have been tested. More pulse shapes should be tested in order to gain a deeper understanding of other pulse characteristics and their effects on the performance of PAPR reduction. For the broadband pulses, truncations of time domain pulses cause some loss in pulse information exhibiting oscillations in the frequency domain. These subsequent oscillation errors also require some further study to quantify their effects on the effectiveness of OFDM transmissions. In the case of PAPR reduction, small frequency oscillations or overshoot errors are relatively negligible for pulses defined over sufficient length durations.